\documentclass[12pt]{article}
\usepackage{epsfig}

\def\beq{\begin{equation}}
\def\eeq#1{\label{#1}\end{equation}}
\def\eeqn{\end{equation}}

\def\beqa{\begin{eqnarray}}
\def\eeqa#1{\label{#1}\end{eqnarray}}
\def\eeqan{\end{eqnarray}}

\let\bar=\overbar

\def\Dslash{\not{\hbox{\kern-4pt $D$}}}
\def\dslash{\not{\hbox{\kern-2pt $\del$}}}

\def\msb{{\bar{\ssstyle M \kern -1pt S}}}

\usepackage{fancyhdr,graphicx}
\def\Title#1{\begin{center} {\Large {\bf #1} } \end{center}}

\begin{document}

\Title{State of matter for quark stars}

\bigskip\bigskip


\begin{raggedright}

{\it Xiaoyu Lai\index{Lai, X. Y.}\\
Department of Astronomy\\
School of Physics\\
Peking University\\
Beijing 100871\\
P. R. China\\
{\tt Email: xylai4861@gmail.com}}
\bigskip\bigskip
\end{raggedright}

\begin{abstract}
It depends on the state of matter at supra-nuclear density to model
pulsar's structure, which is unfortunately not certain due to the
difficulties in physics.
In cold quark matter at realistic baryon densities of compact stars
(with an average value of $\sim 2-3\rho_0$), the interaction between
quarks is so strong that they would condensate in position space to
form quark-clusters.
We argue that quarks in quark stars are grouped in clusters, then we
apply two phenomenological models for quark stars, the polytropic
model and Lennard-Jones model.
Both of the two models have stiffer EoS, and larger maximum mass for
quark stars (larger than 2 $M_\odot$).
The gravitational energy releases during the AIQ process could
explain the observed energy of three supergiant flares from soft
gamma-ray repeaters ($\sim 10^{47}$ ergs).
\end{abstract}

\section{Introduction}

It depends on the state of matter at supra-nuclear density to model
pulsar's structure, which is unfortunately not certain due to the
difficulties in physics although some efforts have been made for
understanding the behavior of quantum chromo-dynamics (QCD) at high
density.
Of particular interest is whether the density in such compact stars
could be high enough to result in unconfined quarks (quark matter).
Stars composed of quarks (and possible gluons) as the dominant
degrees of freedom are called quark stars, and there is possible
observational evidence that pulsar-like stars could be quark
stars(see reviews, e.g.,~\cite{Weber:2004kj,Xu:2008nd}).
But it is still a problem to model a realistic quark star for our
lack of knowledge about the real state of quark matter.

The study of cold quark matter opens a unique window to connect
three active fields: particle physics, condensed matter physics, and
astrophysics.
Many possible states(see, e.g.,~\cite{Alford:2007xm}) of cold quark
matter are proposed in effective QCD models as well as in
phenomenological models.
An interesting suggestion is that quark matter could be in a solid
state~\cite{Xu:2003xe,Horvath:2005ta,Owen:2005fn,Mannarelli:2007bs},
since the strong interaction may render quarks grouped in clusters
and the ground state of realistic quark matter might not be that of
Fermi gas{see a recent discussion given by~\cite{Xu:2008nd}).
If the residua interaction between quark clusters is stronger than
their kinetic energy, each quark cluster could be trapped in the
potential well and cold quark matter will be in a solid state.
Solid quark stars still cannot be ruled out in both astrophysics and
particle physics~\cite{Horvath:2005ta,Owen:2005fn}.
Additionally, there is evidence that the interaction between quarks
is very strong in hot quark-gluon plasma(i.e., the strongly coupled
quark-gluon plasma,~\cite{Shuryak:2008eq}), according to the recent
achievements of relativistic heavy ion collision experiments.
When the temperature goes down, it is reasonable to conjecture that
the interaction between quarks should be stronger than that in the
hot quark-gluon plasma.

Because of the difficulty to obtain a realistic state equation of
cold quark matter at a few nuclear densities, we apply two
phenomenological models, the polytropic model~\cite{Lai:2008cw} and
Lennard-Jones model~\cite{lx09b}, which would have some implications
about the properties of QCD at low energy scale if the astronomical
observations can provide us with some limitations on such models.

This paper is arranged as follows.
The polytropic model is described in Section 2.
The Lennard-Jones model is described in Section 3.
We make conclusions and discussions in Section 4.

\section{The polytropic model}

Because one may draw naively an analogy between the clusters in
quark matter and the nucleus in normal matter, we apply a polytropic
equation of state to quark stars, with different polytropic indices,
$n$.
This model could be regarded as an extension to the quark star
model with a linear equation of state.
We are going to model quark stars in two separated ways.

(i) The vacuum inside and outside of a quark star is assumed the
same, i.e., quark stars have no QCD vacuum energy.
In this case, the equation of state for a quark star is the standard
polytropic model, with a non-zero surface density, representing the
strong confinement between quarks.
Stars of perfect fluid in general relativity was discussed
by~\cite{Tooper}, with an equation of state,
\begin{equation} P=K{\rho_g}^\Gamma, \end{equation}
\begin{equation} \epsilon=\rho_g c^2+n P,
\end{equation}
where $\rho_g$ is that part of the mass density which satisfies a
continuity equation and is therefore conserved throughout the
motion, and $\Gamma=1+1/n$.
Because the quark clusters in quark stars are non-relativistic
particles, the equation of state can be written as
\begin{equation} P=K{\rho_g}^\Gamma, \end{equation}
\begin{equation} \epsilon=\rho_g c^2.
\end{equation}

The mass and radius of a star are evaluated at the point when the
density reaches the surface density, which is non-zero because of
the strong interaction.

(ii) The vacuum energy inside and outside of a quark star are
different.
In this case, the equation of state is \begin{equation} P=K
\rho_g^{1+\frac{1}{n}}-\Lambda, \end{equation} \begin{equation}
\epsilon=\rho_g c^2+\Lambda.
\end{equation}
The density at surface (where pressure is zero) should also be
non-zero.

The key difference between polytropic quark star and normal star
models lies on the surface density $\rho_s$ ($\rho_s
> 0$ for the former but $\rho_s = 0$ for the latter), since a quark
star could be bound not only by gravity but also by additional
strong interaction due to the strong confinement between quarks.
The non-zero surface density is also natural in the case with the
linear equation of state, where the binding effect is represented by
the bag constant, $B$ (and then $\rho_s=4B$).

For stars of perfect fluid, the hydrostatic equilibrium condition
reads
\begin{equation} \frac{1-2GM(r)/c^2r}{P+\rho c^2}r^2 \frac{{\rm d}
P}{{\rm d}r}+\frac{GM(r)}{c^2}+\frac{4\pi G}{c^4}r^3P=0,
\end{equation} where
\begin{equation} M(r)=\int^R_0 \epsilon/c^2 \cdot 4\pi r^2 {\rm d}r.
\end{equation}

For stars with anisotropic pressure, the hydrostatic equilibrium
condition is different.
To simplify the problem, we only consider the case of spherical
symmetry, that the tangential and radial pressure are not equal.
In this case the hydrostatic equilibrium condition
reads(e.g.,~\cite{xty06})
\begin{equation} \frac{1-2GM(r)/c^2r}{P+\rho c^2}(r^2 \frac{d
P}{dr}-2\varepsilon r p)+\frac{GM(r)}{c^2}+\frac{4\pi
G}{c^4}r^3P=0,\end{equation}
where $\varepsilon$ is defined by $P_\perp=(1+\varepsilon)P$, and
$P$ is the radial pressure and $P_\perp$ is the tangential one.
Combine the hydrostatic equilibrium condition and equation of state,
one can calculate the structures of quark stars with and without QCD
vacuum energy.

\subsection{Mass-radius curves}

The mass-radius relation for different polytropic indices, $n$, with
surface density $\rho_s=1.5\rho_0$ ($\rho_0$ is the nuclear matter
density) is shown in Fig~\ref{fig:f1}.
%
\begin{figure}[htb]
\begin{center}
\epsfig{file=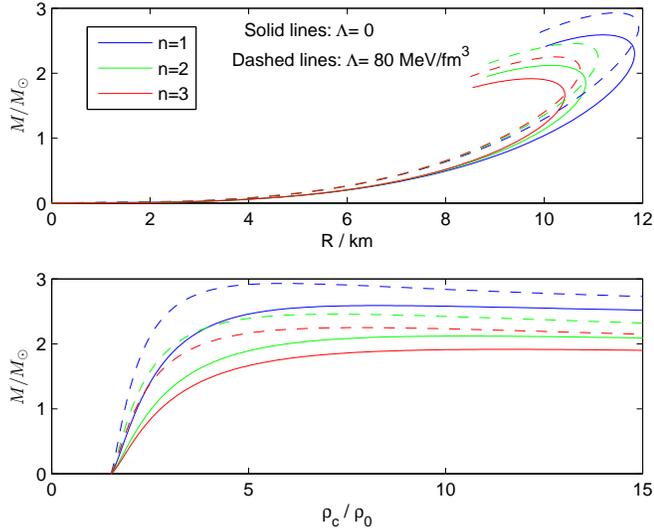,height=3in} \caption{Mass-radius relations for
different polytropic indices, $n$, with $\rho_s=1.5\rho_0$. Solid
lines are $\Lambda=80\rm{MeV fm^{-3}}$, and dash-dotted lines for
$\Lambda=0$. Here and in the following figures, $\rho_0$ is the
nuclear saturation density.} \label{fig:f1}
\end{center}
\end{figure}
%

It is evident from the calculation that the maximum mass of quark
star decreases as the index, $n$, increases.
This is understandable.
A small $n$ means a large $\Gamma$, and the
pressure is relatively lower for higher values of $n$. Lower
pressure should certainly support a lower mass of star.

{\em On the gravitational stability.}
A polytropic star, with a state equation of $P\propto \rho^\Gamma$,
supports itself against gravity by pressure, $PR^2$ (note: the
stellar gravity $\propto M/R^2\propto\rho R$).
Certainly, a high pressure (and thus large $\Gamma$ or small $n$) is
necessary for a gravitationally stable star, otherwise a star could
be unstable due to strong gravity.
Actually, in the Newtonian gravity, a polytropic normal star (with
$\rho_s=0$) is gravitationally unstable if $n>3$, but the star
should be still unstable if $n=3$ when the GR effect is
included~\cite{Shapiro Teukolsky}.

A polytropic quark star with non-zero surface density or with QCD
vacuum energy, however, can still be gravitationally stable even if
$n\geq 3$.
A quark star with much low mass could be self-bound dominantly, and
the gravity is negligible (thus not being gravitationally unstable).
As the stellar mass increases, the gravitational effect becomes more
and more significant, and finally the star could be gravitationally
unstable when the mass increases beyond the maximum mass.
The allowed region for central densities of stable stars are very
narrow ($<\sim 5\rho_0$) for the chosen values of index $n$.

\subsection{Gravitational energy released during a star quake}

The gravitational energy difference between stars with $\varepsilon
\neq 0$ and with $\varepsilon=0$ are shown in Fig~\ref{fig:f2} for
$n=1$.

\begin{figure}[htb]
\begin{center}
\epsfig{file=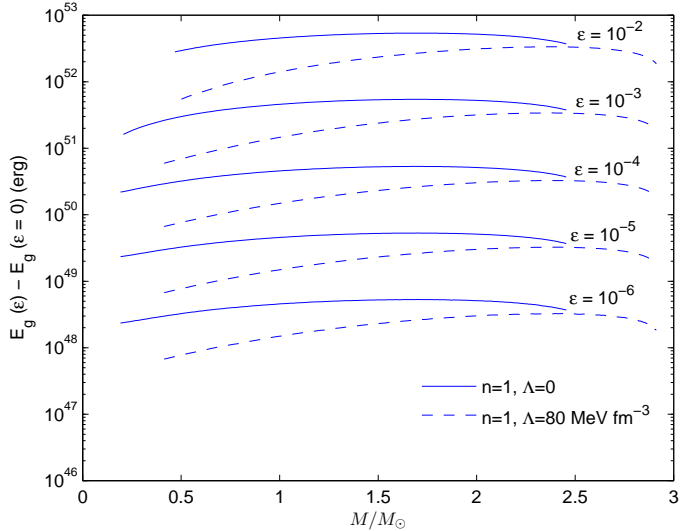,height=3in} \caption{The gravitational energy
difference between stars with and without anisotropic pressures,
which may be released during sequential star quakes.} \label{fig:f2}
\end{center}
\end{figure}
%
Three supergiant flares from soft $\gamma$-ray repeaters have been
observed, with released photon energy being order of $\sim 10^{47}$
ergs.
Our numerical results imply that for all the parameters we chosen,
the released energy could be as high as the observed.

\section{Lennard-Jones model}

The inter-cluster potential could be written as the form of the
potential between two inert gas molecules~\cite{LJ1924}
\begin{equation}u(r)=4U_0[(\frac{r_0}{r})^{12}-(\frac{r_0}{r})^6],\end{equation}
where $U_0$ is the depth of the potential and $r_0$ can be
considered as the range of interaction.
This form of potential has the property of short-distance repulsion
and long-distance attraction.

We suppose that clusters form the simple-cubic structure, and then
from the inter-cluster potential we can get the total energy density
for cold quark matter
\begin{equation}\epsilon_{\rm q}=2U_0(A_{12}r_0^{12}n^5-A_6 r_0^6
n^3)\nonumber+\frac{9}{8}(6\pi^2)^{\frac{1}{3}}\hbar v
n^{\frac{4}{3}}+nm_{\rm c}c^2,\end{equation}
where the first term comes from the potential between clusters, the
second term comes from lattice vibration, and the last term comes
from rest mass energy.
$m_{\rm c}$ is the mass of each quark-cluster, $A_{12}=6.2$ and
$A_6=8.4$.
The pressure can be derived as
\begin{eqnarray}P_{\rm q}&=&n^2\frac{d(\epsilon_{\rm q}/n)}{dn}\nonumber\\
&=&4U_0(2A_{12}r_0^{12}n^5-A_6r_0^6n^3)+\frac{3}{8}(6\pi^2)^{\frac{1}{3}}\hbar
v n^{\frac{4}{3}}.\end{eqnarray}
The model of quark stars composed of Lennard-Jones matter is much
different from the conventional models (e.g., MIT bag model) in
which the ground state is of Fermi gas. In the former case the
quark-clusters are non-relativistic particles, whereas in the the
latter case quarks are relativistic particles.
Consequently, the equations of state in this two kinds of models are
different, and we find that the Lennard-Jones model has some more
stiffer equations of state, which lead to higher maximum masses for
quark stars.
The mass-radius curves and mass-central density curves (the central
density only includes the rest mass energy density), as are shown in
Fig~\ref{fig:f3}.

\begin{figure}[htb]
\begin{center}
\epsfig{file=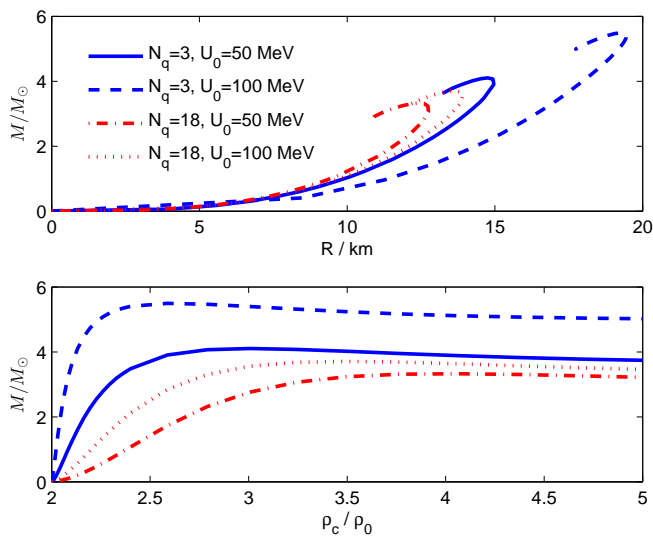,height=3in} \caption{The mass-radius and
mass-central density (rest-mass energy density) curves, in the case
$N_{\rm q}=3$, including $U_0=50$ MeV (blue solid lines) and
$U_0=100$ MeV (blue dashed lines), and the corresponding case
$N_{\rm q}=18$ with $U_0=50$ MeV (red dash-dotted lines) and
$U_0=100$ MeV (red dotted lines), for a given surface density
$\rho_{\rm s}=2\rho_0$.} \label{fig:f3}
\end{center}
\end{figure}

Because of stiffer equations of state, the maximum masses of quark
stars in our model could be higher.
In Fig.2, we can see that (i) a deeper potential well $U_0$ means a
higher maximum mass; (ii) if there are more quarks in a
quark-cluster, the maximum mass of a quark star will be lower.

A stiffer equation of state leading to a higher maximum mass could
have very important astrophysical implications.
Though we have not definitely detected any pulsar whose mass is
higher than 2$M_{\odot}$ up to now, the Lennard-Jones quark star
model could be supported if massive pulsars ($>2M_\odot$) are
discovered in the future.

\emph{Comparison with the MIT bag model.}
In the MIT bag model, quark matter is composed of massless up and
down quarks, massive strange quarks, and few electrons.
Quarks are combined together by an extra pressure, denoted by the
bag constant $B$.
In our model, quarks are grouped in clusters and these clusters are
non-relativistic particles.
If the inter-cluster potential can be described as the Lennard-Jones
form, the equation of state can be very stiff, because at a small
inter-cluster distance (i.e., the number density is large enough),
there is a very strong repulsion.
Whereas in MIT bag model quarks are relativistic particles (at least
for up and down quarks).
For a relativistic system, the pressure is proportional to the
energy density, so it cannot have stiff equation of state.

\section{Conclusions and Discussions}

In cold quark matter at realistic baryon densities of compact stars
(with an average value of $\sim 2-3\rho_0$), the interaction between
quarks is so strong that they would condensate in position space to
form quark-clusters.
Like the classical solid, if the inter-cluster potential is deep
enough to trap the clusters in the potential wells, the quark matter
would crystallize and form solid quark stars.
This picture of quark stars is different from the one in which
quarks form cooper pairs and quark stars are consequently color
super-conductive.

We argue that quarks in quark stars are grouped in clusters, then we
apply two phenomenological models for quark stars, the polytropic
model and Lennard-Jones model.
Both of the two models have stiffer EoS, and larger maximum mass for
quark stars (larger than 2 $M_\odot$).
The gravitational energy releases during the AIQ process could
explain the observed energy of three supergiant flares from soft
gamma-ray repeaters ($\sim 10^{47}$ ergs).

\section*{Acknowledgments}

I am grateful to the members at pulsar group of PKU.


\begin{thebibliography}{99}



\bibitem{Weber:2004kj}
  F.~Weber,
  Prog.\ Part.\ Nucl.\ Phys.\  {\bf 54}, 193 (2005)
  [arXiv:astro-ph/0407155].

\bibitem{Xu:2008nd}
  R.~Xu,
  J. Phys. G, {\bf 36}, 064010 (2009)
  [arXiv:0812.4491 [astro-ph]].

\bibitem{Alford:2007xm}
  M.~G.~Alford, A.~Schmitt, K.~Rajagopal and T.~Schafer,
  Rev.\ Mod.\ Phys.\  {\bf 80}, 1455 (2008)
  [arXiv:0709.4635 [hep-ph]].

\bibitem{Xu:2003xe}
  R.~X.~Xu,
  Astrophys.\ J.\  {\bf 596}, L59 (2003)
  [arXiv:astro-ph/0302165].

\bibitem{Horvath:2005ta}
  J.~E.~Horvath,
  Mod.\ Phys.\ Lett.\  A {\bf 20}, 2799 (2005)
  [arXiv:astro-ph/0508223].

\bibitem{Owen:2005fn}
  B.~J.~Owen,
  Phys.\ Rev.\ Lett.\  {\bf 95}, 211101 (2005)
  [arXiv:astro-ph/0503399].

\bibitem{Mannarelli:2007bs}
  M.~Mannarelli, K.~Rajagopal and R.~Sharma,
  Phys.\ Rev.\  D {\bf 76}, 074026 (2007)
  [arXiv:hep-ph/0702021].

\bibitem{Shuryak:2008eq}
  E.~Shuryak,
  Prog.\ Part.\ Nucl.\ Phys.\  {\bf 62}, 48 (2009)
  [arXiv:0807.3033 [hep-ph]].

\bibitem{Lai:2008cw}
  X.~Y.~Lai and R.~X.~Xu,
  Astropart.\ Phys.\  {\bf 31}, 128 (2009)
  [arXiv:0804.0983 [astro-ph]].



\bibitem{lx09b}
X. Y. Lai and R. X. Xu, MNRAS, {\bf 398}, L31 (2009)


\bibitem{Shapiro Teukolsky}
S. L. Shapiro and S. A. Teukolsky, {\it Black Holes, White Dwarfs,
and Neutron Stars: the physics of compact objects} (New York:
Wiley-Interscience, 1983)

\bibitem{Tooper}
R. F. Tooper, ApJ, {\bf 142}, 1541 (1965).

\bibitem{xty06}
R. X. Xu, D. J. Tao and Y. Yang, Mon. Not. R. Astron. Soc., {\bf
373}, L85 (2006).

\bibitem{LJ1924}
J. E. Lennard-Jones, Proc. Roy. Soc. (London) A{\bf 106}, 463
(1924).




\end{thebibliography}
\end{document}